# Current progress in vanadium oxide nanostructures and its composites as supercapacitor electrodes

Raktima Basu,* Sandip Dhara*

Surface and Nanoscience Division, Indira Gandhi Centre for Atomic Research, Homi Bhabha National Institute, Kalpakkam-603102, India

Email: raktimabasu14@gmail.com;dhara@igcar.gov.in

## *Abstract*

In recent years, vanadium oxides have gained immense attention in the field of energy storage devices due to their low-cost, layered structure and multi-valency despite their limited electrical conductivity and lower structural stability. In this brief review, we have tried to focus on electrochemical properties of the stoichiometric vanadium oxides along with VO_x composites. The morphology engineering, doping with heteroatom and formation of composites with carbon-based materials and/or conducting polymers in enhancing the supercapacitive performances of the vanadium oxides are discussed in details. Finally, the potentiality and challenges of vanadium oxides nanocomposites for supercapacitor applications are discussed.

**Keywords:** supercapacitor, vanadium oxide, specific capacitance, pseudocapacitance, phase transition

## 1. Introduction:

In recent time, supercapacitors (SCs) are one of the emerging technologies used for clean energy prospect. The higher power density, low specific energy, longer cycle life, and environmental affability made the SCs superior compared to conventional batteries. However, the scientific community is working towards increasing the specific energy of SCs by finding a suitable electrode material. Carbon materials, conducting polymers, and metal oxide or hydroxides are reported to be suitable candidates as electrodes for SC [1-3]. Carbon materials such as activated carbon, carbon nanotube and graphene provide excellent electrical conductivity and chemical stability [4], however, they come with narrow charge storage capacity and relatively low energy density [1]. On the other hand, the conducting polymers are a good choice as a pseudocapacitor [3]. Nevertheless, the electrochemical stability is poor. Towards this end, transition metal oxides (TMOs) are alternative candidates due to their multiple oxidation states and rapid redox kinetics [2]. Amongst other TMOs [5-7], vanadium oxides have received recent attention owing to their low-cost, variety of valence states, and abundant sources [8-10]. V is a transition metal ([Ar]$3d^3 4s^2$) with valences in the range of +2 to +5 with major oxides as VO, $V_2O_3$, $VO_2$, and $V_2O_5$ [11]. However, the V-O phase diagram comprises mixed-valence oxides comprehending two oxidation states, e.g. $V_4O_9$, $V_6O_{13}$, $V_8O_{15}$, $V_7O_{13}$, $V_6O_{11}$, among others which permit conversion between oxides of different stoichiometry easily and make it unstable during charging- discharging cycle. As the composition, oxidization state, and structural phase of a material have a significant role on electrochemical properties; the exchange of valency along with structural instability for these materials results in poor electrochemical and cycle performance. Issues related to the presence of multiple valence states of V [10], as well as its stability affecting retention of capacitance and its efficiency, are found to hinder further utility in SCs. The poor charge storage properties and electrical conductivity of vanadium oxides are reported to be succeeded by fabricating directly on the current collector, doping element, or by nanostructure engineering. There are several reports on SC properties of $VO_x$ composites as well as individual vanadium oxides such as $VO_2$, $V_2O_3$, and $V_2O_5$ [3, 12-23]. In this context, this mini review presents a summary of recent developments in vanadium oxide based supercapacitor along with future developments, prospects, and challenges.

## 2. Electrochemical properties of vanadium dioxide:

Vanadium dioxide ($VO_2$) is known to be stabilized in different polymorphs, including $VO_2$(A), $VO_2$(B), $VO_2$(C), $VO_2$(D), among others. [24]. Among the $VO_2$ polymorphs, $VO_2$(B) attracted much attention for its well known MIT at a technologically important temperature of 340K, which is very close to room temperature [25]. $VO_2$(B) crystallizes in rutile tetragonal (R; space group $P4_2/mnm$) and monoclinic (M1; space group $P2_1/c$) structure above and below the transition temperature, respectively [26,27]. In the high-temperature R phase, V atoms are equally spaced, forming linear chains along the $c_R$ axis with each V atom surrounded by an oxygen octahedron [28]. The lattice parameters are $c_R$ = 2.85 Å, and $a_R = b_R$ = 4.55 Å. Whereas in the low-temperature monoclinic phase the volume of the unit cell becomes double than that of R phase with lattice parameters $a_{M1}$ = 5.70 Å, $b_{M1}$ = 4.55 Å, $c_{M1}$ = 5.38 Å, and $ß_{M1}$ = 123° [29]. The approximate crystallographic relationship between M1 and R phase is $a_{M1} \leftrightarrow 2c_R$, $b_{M1} \leftrightarrow a_R$, and $c_{M1} \leftrightarrow b_R - c_R$ [30]. In the M1 phase, there are significant differences in the arrangement of V along $c_R$ axis. The V atoms form pair, and the pairs tilt along the $c_R$ axis making the surrounding octahedron deformed. Besides M1, two more metastable phases of monoclinic M2 having space group $C2/m$ and triclinic T (alternatively monoclinic M3) with space group $\bar{C}1$ are also reported in the process of the phase transition from M1 to R [31].

There are several reports on the supercapacitive performance of $VO_2$(B) in the M1 phase. However, its low rate capability and cycling instability become the obstacles to serve as a commercial supercapacitor. The modification in structure designing has been adopted to overcome the barrier. Zhang *et al*. [32] prepared a template-free 3D hollow spherical cages (shown in figures 1a-b) by hydrothermal method, which showed a specific capacitance of 1175 $mF\cdot cm^{-2}$ (336 $F\cdot g^{-1}$) with enhanced stability, and 68% of the capacitance was retained after 10000 cycles. However, 2D nanosheet of $VO_2$ is reported to be a more eligible candidate for electrochemical performance than that of 3D counterpart because of large specific surface area shortening the diffusion path of ion and enhance the redox reaction. Ndiaye *et al*. [33] obtained specific capacitance of 663 $F\cdot g^{-1}$ at the scan rate of 5 $mV\cdot s^{-1}$ and excellent cycling stability after 5000 cycles at the current density of 10 $A\cdot g^{-1}$ for $VO_2$ nanosheets. The 2D nanosheets (figures 1c-d), while assembled with the structure of carbonized iron-polyaniline (C-FP), exhibited a specific capacity of 47 $mAh\cdot g^{-1}$ at a current density of 1 $A\cdot g^{-1}$ [34]. In 2D nanosheets, a large specific surface area diminishes the path length of the ion diffusion enabling the execution of the redox reaction effectively. Rakhi *et al*. [35] reported a specific capacitance of ~ 405 $F\cdot g^{-1}$ at the current density of 1 $A\cdot g^{-1}$ of $VO_2$ nanosheets in an organic gel electrolyte (1 M $LiClO_4$ in propylene carbonate) with nearly 82% capacitance retention.

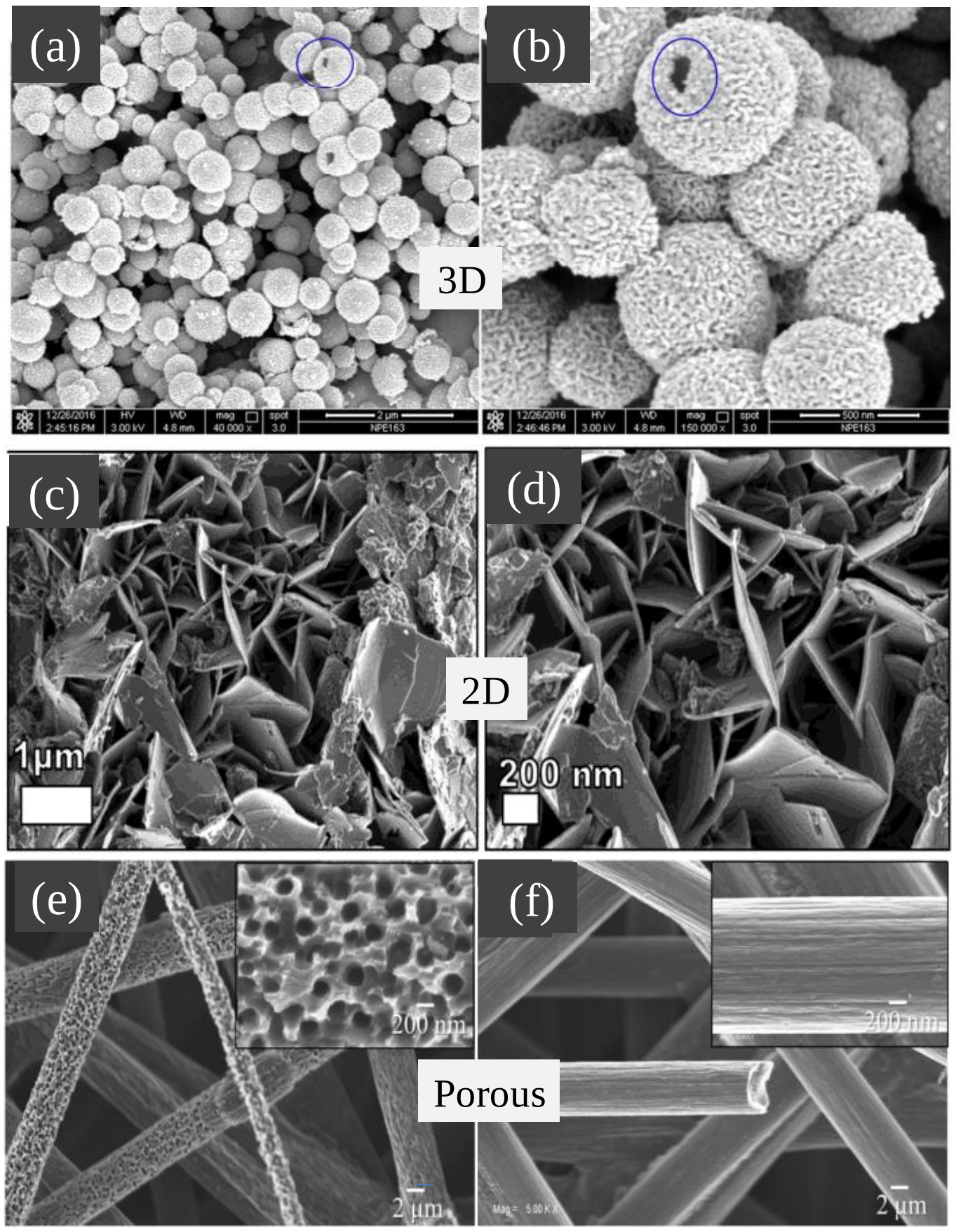

Figure 1. Scanning electron micrograph of (a-b) $VO_2$ 3D hollow spherical cages and its higher magnification image; marked circles shows broken area [*Reprinted with permission from Ref. 33, Copyright 2018 @ Royal Society of Chemistry*] (c-d) 2D $VO_2$ nanosheets and its zoomed image [*Reprinted with permission of authors from Ref. 34, Copyright 2019 @ American Institute of Physics*], (e) $VO_2$ nanoporous structure grown on carbon paper, (f) bare C fiber [*Reprinted with permission of authors from Ref. 37, Copyright 2019 @ Nature Publishing Group*].

The thin layer of 1D $VO_2$ nanorods on indium tin oxide - coated glass substrates are also reported [36] to produce a specific capacitance of ~486 $mF{\cdot}cm^{-2}$ at the scan rate of 10 $mV{\cdot}s^{-1}$. Nie *et al*. [19] reported $VO_2$@Polyaniline coaxial nanobelts exhibiting a higher specific capacitance of 246 $F{\cdot}g^{-1}$ at 0.5 A $g^{-1}$ than that of $VO_2$ nanobelts (160.9 $F{\cdot}g^{-1}$). The specific capacitance was almost constant at around 118 $mF{\cdot}cm^{-2}$ after 5000 cycles at the scan rate of 100 $mV{\cdot}s^{-1}$. $VO_2$ nanoporous structures on carbon fiber in the M1 phase (figures 1e-f) exhibit a specific capacitance of 20.7

mF·cm$^{-2}$ at the current density of 0.3 mA·cm$^{-2}$ [37]. It also demonstrates capacitance retention of 93.7% and coulombic efficiency of 98.2% for 5000 charge-discharge cycles. However, the similar nanoporous structures in M2 and T phases of $VO_2$ show poor specific capacitance (figure 2a) as well as cyclic stability (figure 2b) because of mixed valency [37].

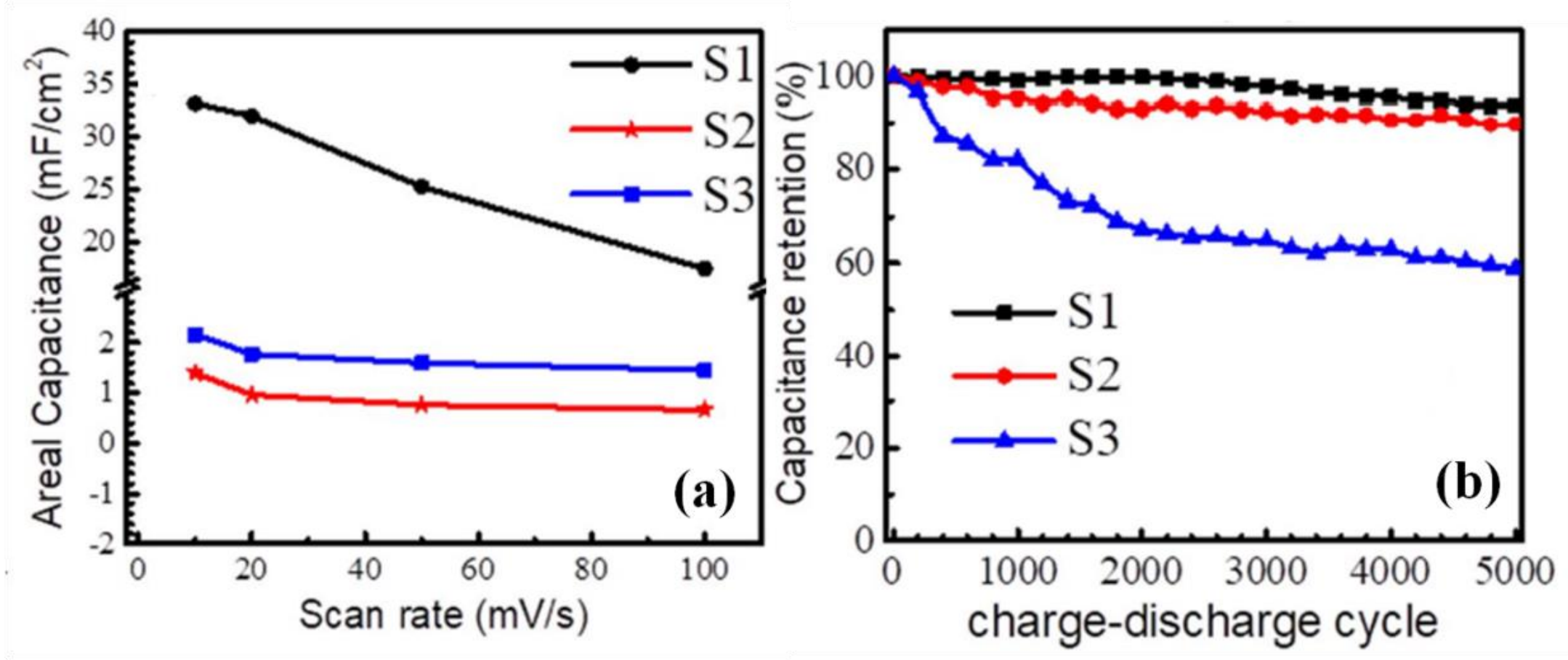

Figure 2. (a) Areal capacitance versus scan rate and (b) capacitance retention with charge-discharge cycle of pure M1 (S1), and mixed phases of $VO_2$ (S2 and S3) [*Reprinted with permission of authors from Ref. 37, Copyright 2019 @ Nature Publishing Group*].

Another way to enhance the electrochemical performance of $VO_2$ is by combining with carbon materials, which improve the electrical conductivity. $VO_2$ nanoflowers on 3D graphene networks were reported to exhibit a large specific capacitance of 466 mF·cm$^{-2}$, capacitance retention of 63.5% after 3000 cycles by Wang *et al.* [38]. Ren *et al.* [39] synthesized $VO_2$ nanoparticles on edge-oriented graphene foam (EOGF) which exhibit a capacitance of 119 mF·cm$^{-2}$ at the scan rate of 2 mV·s$^{-1}$. The $VO_2$(B)/carbon core-shell composites prepared by Zhang *et al.* [40] exhibited a specific capacitance of 203 F·g$^{-1}$ at the current density of 0.2 A·g$^{-1}$. Lv *et al.* [41] prepared $VO_2$(B) nanobelts/rGO composites with a porous framework, which showed an excellent power density of 7152 W·kg$^{-1}$ at the energy density of 3.13 Wh·kg$^{-1}$. Shao *et al*. [20] reported $VO_2$ demonstrating superior properties as supercapacitor compared to that for the $V_2O_5$, which was well known for its SC performance. It is due to higher electronic conductivity in $VO_2$, as compared to $V_2O_5$, originating from a mixed-valence and structural stability because of the increased edge sharing and the consequent resistance to lattice shearing during cycling [42]. The comparison of various $VO_2$ based supercapacitors and their synthesis procedures are shown in table 1.

## 3. Electrochemical properties of vanadium trioxide:

Vanadium trioxide ($V_2O_3$) revels a rhombohedral corundum structure at room temperature, (space group $R\overline{3}c$) [43], where the V atoms pair along the crystal *c*-axis and form honeycomb lattices in the *ab*-plane. Whereas, below the temperature ~150 K, a paramagnetic metallic to an antiferro-magnetic insulating transition happens along with the structural transition to the monoclinic phase (space group $I_2/a$) [44].

There are very few reports on the electrochemical studies on $V_2O_3$, mostly because of the poor stability of this material. A binder-free electrode of $V_2O_3$ nanoflakes on N-doped rGO (figure 3a) was reported to have an areal capacitance of 216 $mF{\cdot}cm^{-2}$ at a current density of 1 $mA{\cdot}cm^{-2}$ (figure 3b). It also exhibits cycling stability with retention of ~81% of the initial capacitance value after 10000 cycles (figure 3c) [45].

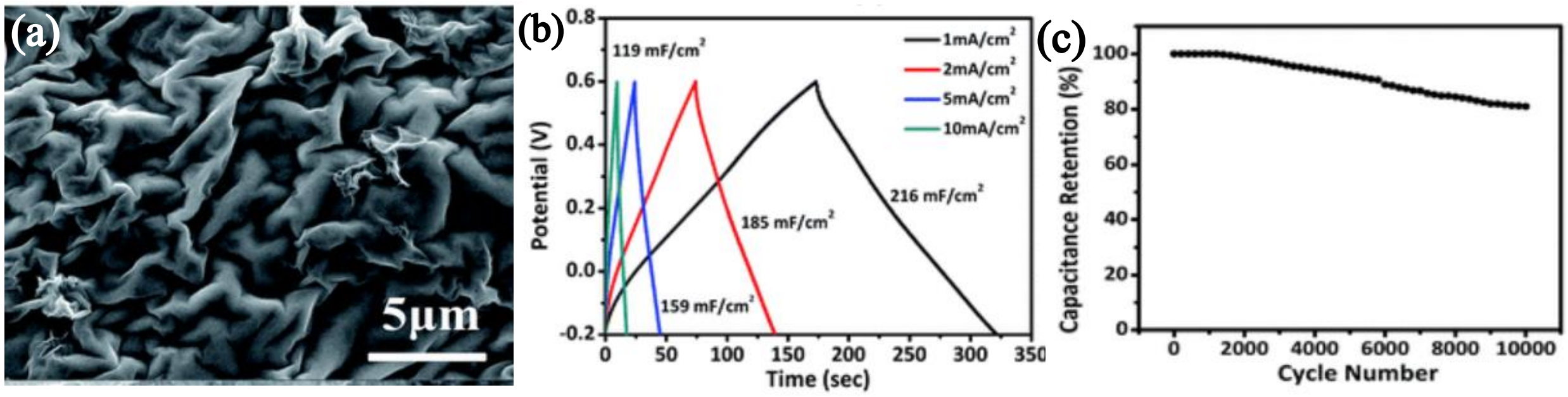

Figure 3. (a) Scanning electron micrograph of the $V_2O_3$/N-rGO samples, (b) The galvanostatic charge-discharge (GCD) curves obtained for the self-supported $V_2O_3$/N-rGO film electrodes at different current densities. (c) Cycling stability for 10000 cycles [*Reprinted with permission from Ref. 45, Copyright 2017 @ Royal Society of Chemistry*].

Table1: Comparison of various $VO_2$ based supercapacitors

| **Nanostructures (Growth Technique)** | **Electrolyte** | **Specific Capacitance** | **Current density** | **Cycling stability (%)** | **Ref.** |
|---|---|---|---|---|---|
| $VO_2$ (B) hollow spheres (Solvothermal) | 1 M $Na_2SO_4$/ critical micelle concentration | 336 $F·g^{-1}$ | 2 $mA·cm^{-2}$ | 68% (10000 cycles) | 33 |
| $VO_2$ nanosheets (Solvothermal) | 6 M KOH | 663 $F·g^{-1}$ | 10 $A·g^{-1}$ | 99.4% (9000 cycles) | 34 |
| $VO_2$ nanosheets (Solvothermal) | 6 M KOH | 47 $mAh·g^{-1}$ | 1 $A·g^{-1}$ | 89% (10 000 cycles) | 34 |
| $VO_2$ nanosheet (Solution Reduction of hydrothermally exfoliated bulk $V_2O_5$) | 1 M $LiClO_4$/PPC | 405 $F·g^{-1}$ | 1 $A·g^{-1}$ | 82% (6000 cycles) | 35 |
| $VO_2$ nanorod thin films (RF magnetron sputtering) | 0.1 M NaOH | 486 $mF·cm^{-2}$ | 10 $mV·s^{-1}$ | 100% (5000 cycles) | 36 |
| $VO_2$@Polyaniline coaxial nanobelts (Reactive templated organic layer on solvothermally grown $VO_2$ nanobelt) | 0.5 M $Na_2SO_4$ | 246 F $g^{-1}$· | 0.5 A $g^{-1}$ | (28.6%) (1000 cycles) | 19 |
| Nanoporous $VO_2$ (Vapour transport of bulk $V_2O_5$ on C-paper) | $Na_2SO_4$ | 20.7 $mF·cm^{-2}$ | 0.3 $mA·cm^{-2}$ | 93.7% (5000 cycles) | 37 |
| $VO_2$ NFs@3DG (Hydrothermally grown $VO_2$ on 3DG) | 0.5 M $K_2SO_4$ | 507 $F·g^{-1}$ | 3 $mA·cm^{-2}$ | 63.5% (3000 cycles) | 38 |
| $VO_2$ nanoparticle/EOGF (Hydrothermally grown $VO_2$ on EOGF) | 5 M LiCl | 119 $mF·cm^{-2}$ | 2 $mV·s^{-1}$ | 70% (1500 cycles) | 39 |
| $VO_2$(B)/C core-shell (Single pot hydrothermal) | 1 M $Na_2SO_4$ | 203 $F·g^{-1}$ | 0.2 $A·g^{-1}$ | 10.4% (100 cycles) | 40 |
| $VO_2$(B) nanobelts/rGO (Hydrothermally grown $VO_2$ on rGO) | 0.5 M $K_2SO_4$ | 353 $F·g^{-1}$ | 1 $A·g^{-1}$ | 78% (10 000 cycles) | 41 |

However, $V_2O_3$ combined with carbon composites are reported to serve as superior electrode material. Zheng *et al.* reported $V_2O_3$/C composites exhibiting high pseudocapacitance of 458.6 $F\cdot g^{-1}$ at 0.5 $A\cdot g^{-1}$. The composite also shows a retention rate of 86% after 1000 cycles in aqueous electrolyte [46]. Hu *et al*. [47] synthesized $V_2O_3$@C core-shell nanorods with porous structures which exhibited 228, 221, 207, 158, and 127 $F\cdot g^{-1}$ specific capacitances at current densities of 0.5, 1, 2, 5, and 10 $A\cdot g^{-1}$, respectively. Zhang *et al.* [48] reported a $V_2O_3$ nanofoam@activated carbon composite, which showed a specific capacitance of 185 $F\cdot g^{-1}$ at 0.05 $A\cdot g^{-1}$. The comparison of various $V_2O_3$ based supercapacitors, fabricated following different process steps, are shown in table 2.

Table 2: Comparison of various $V_2O_3$ based supercapacitors

| **Nanostructures (Growth Technique)** | **Electrolyte** | **Specific Capacitance** | **Current density** | **Cycling stability (%)** | **Ref.** |
|---|---|---|---|---|---|
| $V_2O_3$/N-rGO nanoflakes (500 °C $NH_3$ reduction of $V_2O_5$ gel/GO films) | 1 M $Na_2SO_4$ | 216 $mF\cdot cm^{-2}$ | 1 $mA\cdot cm^{-2}$ | 81% (10000 cycles) | 45 |
| $V_2O_3$/C nanocomposites (Calcination of hydrothermally grown $(NH_4)_2V_3O_8$) | 5 M LiCl | 458.6 $F\cdot g^{-1}$ | 0.5 $A\cdot g^{-1}$ | 86% (1000 cycles) | 46 |
| $V_2O_3$@C core-shell nanorods (Single pot hydrothermal process using $V_2O_5$ nanorod) | 5 M LiCl | 228 $F\cdot g^{-1}$ | 0.5 $A\cdot g^{-1}$ | 86% (1000 cycles) | 47 |
| $V_2O_3$ nanofoam@activated carbon (Calcination of $NH_4VO_3$ solution and activated C) | 1 M $NaNO_3$ | 185 $F\cdot g^{-1}$ | 0.05 $A\cdot g^{-1}$ | 49% (100 cycles) | 48 |

## 4. Electrochemical properties of vanadium pentoxide:

Vanadium pentoxide ($V_2O_5$) stabilizes in various phases including $\alpha$-$V_2O_5$, $\beta$-$V_2O_5$, $\delta$-$V_2O_5$, $\gamma'$-$V_2O_5$, $\zeta$-$V_2O_5$, and $\varepsilon'$-$V_2O_5$ [49]. The most well-known phase is $\alpha$-$V_2O_5$, which crystallizes into an orthorhombic structure composed of weakly Van der Walls bonded layers of $VO_5$ pyramids sharing their vertices and corners [50,51]. The unit-cell parameters are $a$ = 11.51 Å, $b$ = 3.56 Å, and $c$ = 4.37 Å [50]. It has space group *Pmmn*, ($D_{2h}^{13}$) with distorted square-pyramidal coordination symmetry around each V atom. There are three non-equivalent oxygen atoms in each unit cell (denoted as $O_I$, $O_{II}$, and $O_{III}$). $O_I$ is the terminal (vanadyl) oxygen with two different bond lengths. One of them is a strong and short V-$O_I$ bond with a length of 1.577 Å ($d_1$). Another one is large and weak Van der Waals type connecting two adjacent layers in the $V_2O_5$ structure, with a bond length of 2.793 Å. Both of these $O_I$ atoms orient almost along the $c$-axis. The two-fold coordinated bridging oxygen ($O_{II}$) connects two adjacent V atoms with V-$O_{II}$ bond length of 1.78 Å ($d_2$). The ladder-shaped $O_{III}$ atoms are the three-fold coordinated oxygen with three different V-$O_{III}$ bond lengths of 1.88 ($d_3$), 1.88 ($d_3$), and 2.02 Å ($d_4$) [50].

The SC properties in $V_2O_5$ is reported to be superior to other vanadium oxides because of its stability and layered structure [16,18,19,20]. Yang *et al*. [52] prepared hollow $V_2O_5$ spheres which showed an excellent capacitance of 479 $F{\cdot}g^{-1}$ at 5 $mV{\cdot}s^{-1}$. $V_2O_5$ nanofibers showed specific capacitance of 190 $F{\cdot}g^{-1}$ in aqueous electrolyte (KCl) and 250 $F{\cdot}g^{-1}$ in the organic electrolyte ($LiClO_4$ in PPC) as reported by Wee *et al*. [53]. Apart from supercapacitor performance, the change in electrolytes in case of $V_2O_5$ also controls its mechanical stability and chemical dissolution. Pandit *et al*. [54] synthesized $V_2O_5$ thin film on a pliable stainless steel substrate which is reported to exhibit a high specific capacitance of 735 $F{\cdot}g^{-1}$ at 1 $mV{\cdot}s^{-1}$ with capacitors retention of 71% after 1000 cycles.

The rGO/$V_2O_5$ composites showed specific capacitance of 386, 338, 294, 241, and 197 $F{\cdot}g^{-1}$ at current density of 0.1, 0.2, 0.5, 1, and 2 $A{\cdot}g^{-1}$, respectively, as reported by Liu *et al*. [55]. However, 2D heterostructures of $V_2O_5$ nanosheets growing on rGO flakes showed relatively high specific capacitance of 653 $F{\cdot}g^{-1}$ at 1 $A{\cdot}g^{-1}$ and cyclic stability of 94% after 3000 cycles [56]. Choudhury *et al.* [57] prepared $V_2O_5$ nanofiber (VNF)/exfoliated graphene nanohybrid with the mass ratio of 1:0.25 and 1:0.5 with a superior capacitance value of 218 $F \cdot g^{-1}$ at 1 $A \cdot g^{-1}$ for 1:0.5 mass ratio. Balasubramanian *et al.* [58] reported flowery $V_2O_5$ structures coated with carbon showing specific capacitance of 417 $F \cdot g^{-1}$ at a current density of 0.5 $A \cdot g^{-1}$. Chen *et al.* [59] synthesized $V_2O_5$ nanocomposites with carbon nanotubes (CNT) which provided a capacity of 228 $C \cdot g^{-1}$ between 1.8 and 4.0 V. Wu *et al*. [60] reported $V_2O_5$/multi-walled CNT core/shell hybrid aerogel (figure 4a), which demonstrated the maximum specific capacitance of 625 $F \cdot g^{-1}$ with outstanding cycle performance (>20000 cycles). The hybrid aerogel showed better performance than that of raw $V_2O_5$ powder, MWCNTs, and $V_2O_5$ aerogel (figure 4b).

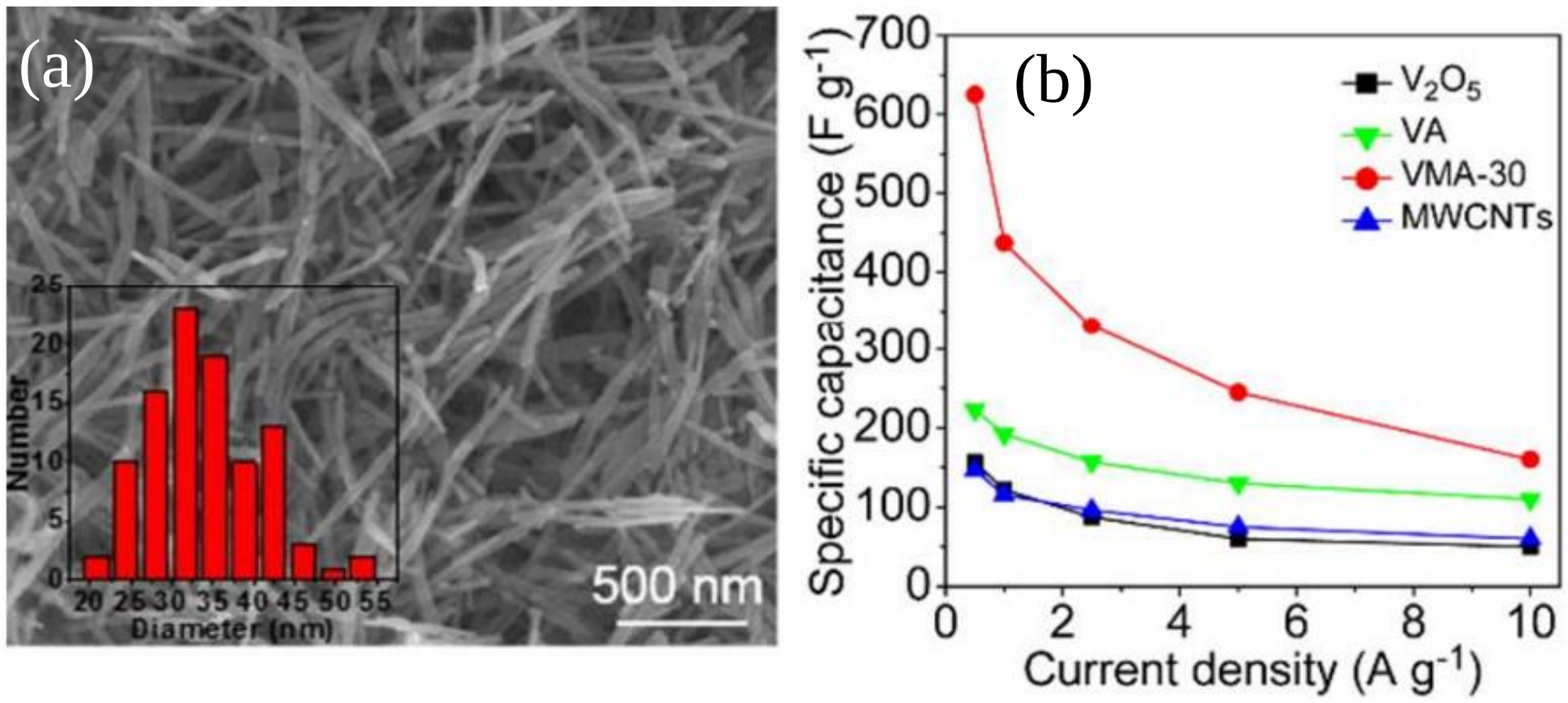

Figure 4. (a) Scanning electron micrograph of $V_2O_5$ aerogel and diameter distribution of $V_2O_5$ nanofibers (inset) (b) Specific capacitance as a function of current density for raw $V_2O_5$ powder, MWCNTs, $V_2O_5$ aerogel and hybrid aerogel of $V_2O_3$/C nanocomposites (VMA-30) [*Reprinted with permission from Ref. 60, Copyright 2015 @ Royal Society of Chemistry*].

However, the insertion of nitrogen atoms into the carbon network enhances the electrochemical performance by restraining the hydrophobicity. Sun *et al.* [61] reported self-assembled 3D N-carbon nanofibers (CNFs)/$V_2O_5$ aerogels showing the specific capacitance of 575.6 $F \cdot g^{-1}$ even after 12000 cycles (97% of the initial value). $V_2O_5$ also have been combined with conducting polymers *e.g.*, polypyrrole (PPy), poly (3, 4-ethylenedioxythiophene) (PEDOT), and polyaniline (PANI), to enhance the electrical conductivity and prevent the V from dissolving. Qian *et al*. [62] reported 3D $V_2O_5$/PPy nanostructures, which exhibited a high specific capacitance of 448 F·g$^{-1}$. However, Bi *et al.* [63] showed a comparative study with oxygen vacancy (Ö) resulting with the specific capacitance of 614 F·g$^{-1}$ for VÖ-$V_2O_5$/PEDOT higher than that of 523 F·g$^{-1}$ for VÖ-$V_2O_5$/PANI and 437 F·g$^{-1}$ for VÖ-$V_2O_5$/PPy (figures 5a-b).

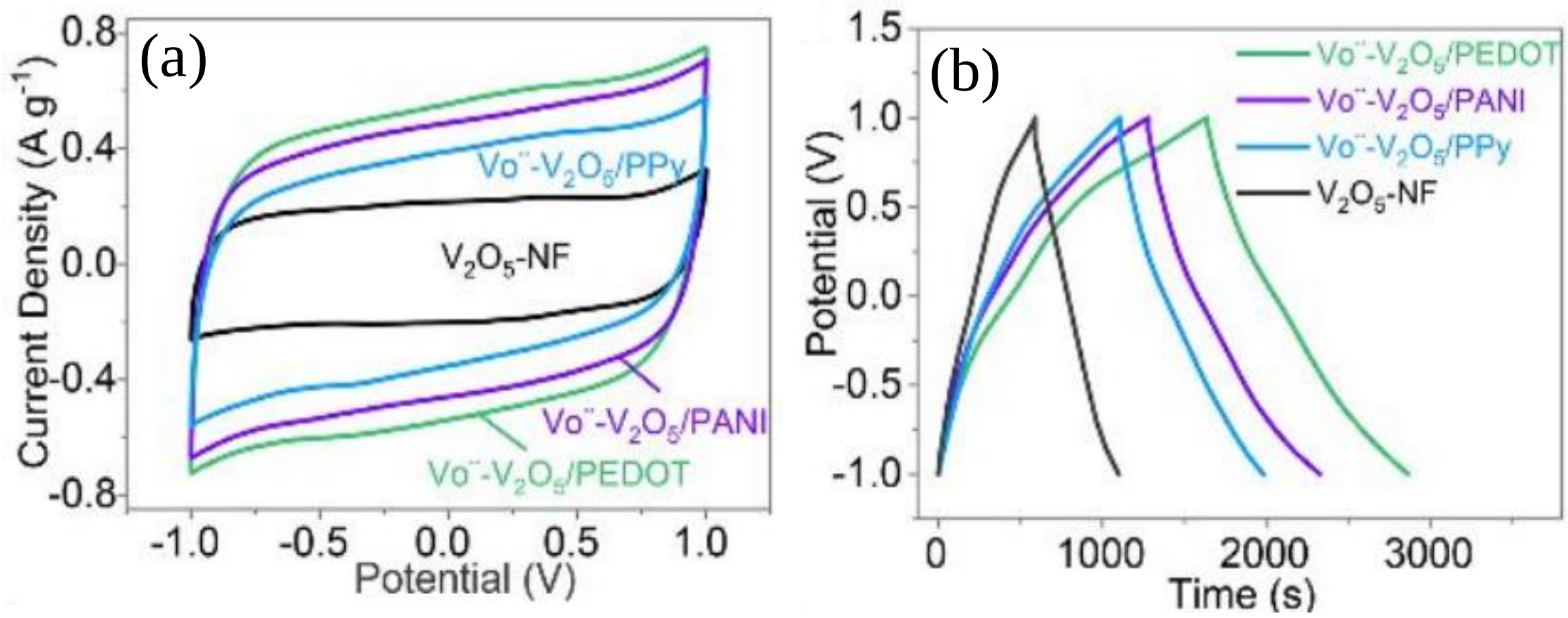

Figure 5. (a) CV curves of V VÖ -$V_2O_5$/Conducting polymers and $V_2O_5$-NF at a scan rate of 5 mV s$^{-1}$. (b) galvanostatic charge-discharge curves of VÖ-$V_2O_5$/ Conducting polymers and $V_2O_5$-NF at a current density of 0.5 A g$^{-1}$ [*Reprinted with permission from Ref. 63, Copyright 2019 @ Royal Society of Chemistry*].

The association of $V_2O_5$ with other metal oxides is also reported to enhance the electrochemical properties. Xu *et al*. prepared $V_2O_5$ nanobelts/$TiO_2$ nanoflakes composites [64], which exhibited high specific capacitance of 587 F·g$^{-1}$ at 0.5 A·g$^{-1}$ with good cyclic stability of 97% after 5000 cycles. The $V_2O_5$-doped α-$Fe_2O_3$ composites by Nie *et al*. [65] showed a capacitance value of 150 F·g$^{-1}$ over 200 cycles at 6 A·g$^{-1}$. CNT-$SnO_2$-$V_2O_5$ composites exhibited higher specific capacitance compared to CNT, $V_2O_5$, and CNT-$V_2O_5$ [66]. $V_2O_5$ shows higher performance combined with carbon and other metal oxides than that of the other two phases of vanadium oxides discussed before. The comparison of various $V_2O_5$ based supercapacitors with involved synthesis techniques are shown in table 3.

Table 3: Comparison of various $V_2O_5$ based supercapacitors

| Nanostructures (Growth Technique) | Electrolyte | Specific Capacitance | Current density | Cycling stability (%) | Ref. |
|---|---|---|---|---|---|
| Hollow spherical $V_2O_5$ (Solvothermal) | 5 M $LiNO_3$ | 479 $F{\cdot}g^{-1}$ | 5 $mV{\cdot}s^{-1}$ | 43% (100 cycles) | 52 |
| $V_2O_5$ nanofibers (Electrospinning) | (i) 2 M KCl (ii) 1 M $LiClO_4$/PPC | 190 $F{\cdot}g^{-1}$ 250 $F{\cdot}g^{-1}$ | 0.1 $A{\cdot}g^{-1}$ | | 53 |
| $V_2O_5$ complex (Chemical bath deposition) | 2 M $LiClO_4$ | 735 $F{\cdot}g^{-1}$ | 1 $mV{\cdot}s^{-1}$ | 71% (1000 cycles) | 54 |
| rGO/$V_2O_5$ hybrid aerogel (One-pot Hydrothermally grown) | 1 M $LiClO_4$/PPC | 384 $F{\cdot}g^{-1}$ | 0.1 $A{\cdot}g^{-1}$ | 82.2% (10 000 cycles) | 55 |
| rGO/$V_2O_5$ nanosheet (Mixing rGO with hydrothermally grown $V_2O_5$ ) | 1 M KCl | 653 $F{\cdot}g^{-1}$ | 1 $A{\cdot}g^{-1}$ | 94% (3000 cycles) | 56 |
| VNF/graphene nanohybrid (Hydrothermally grown VNF mixed with exfoliated graphene) | 1 M LiTFSI in acetonitrile | 218 $F{\cdot}g^{-1}$ | 1 $A{\cdot}g^{-1}$ | 87% (700 cycles) | 57 |
| Carbon coated flower $V_2O_5$ (Co-precipitation method followed by annealing at 400 °C) | 1 M $K_2SO_4$ | 417 $F{\cdot}g^{-1}$ | 0.5 $A{\cdot}g^{-1}$ | 100% (2000 cycles) | 58 |
| CNT/$V_2O_5$ nanocomposite (One-pot hydrothermal process of $V_2O_5$ and hydrophilic CNTs) | 1 M $LiClO_4$/PPC | 228 $C{\cdot}g^{-1}$ | 20 $mV{\cdot}s^{-1}$ | 80% (10000 cycles) | 59 |
| $V_2O_5$/MWCNT core/shell hybrid aerogels (One-step sol-gel process) | 1 M $Na_2SO_4$ | 625 $F{\cdot}g^{-1}$ | 0.5 $A{\cdot}g^{-1}$ | 120% (20 000 cycles) | 60 |
| 3D N-CNFs/$V_2O_5$ aerogels (Self-assembly of nanostructured $V_2O_5$ onto CNF aerogels with N) | 1 M $Na_2SO_4$ | 595.1 $F{\cdot}g^{-1}$ | 0.5 $A{\cdot}g^{-1}$ | 97% (12000 cycles) | 61 |
| 3D $V_2O_5$/PPy core/shell nanostructures ($V_2O_5$ by ion exchange attached with PPy) | 5 M $LiNO_3$ | 448 $F{\cdot}g^{-1}$ | 0.5 $A{\cdot}g^{-1}$ | 81% (1000 cycles) | 62 |
| $V_2O_5$-Conductive polymer nanocables ($V_2O_5$ sol attached to functionalized polymers) | 1 M $Na_2SO_4$ | 614 $F{\cdot}g^{-1}$ | 0.5 $A{\cdot}g^{-1}$ | 111% (15 000 cycles) | 63 |
| $V_2O_5$/$TiO_2$ composites (Two-step hydrothermal process using Ni foam) | 1 M $LiNO_3$ | 587 $F{\cdot}g^{-1}$ | 0.5 $A{\cdot}g^{-1}$ | 92% (1000 cycles) | 64 |
| $V_2O_5$-α-$Fe_2O_3$ composite nanotubes (One-step electrospinning) | 3 M KOH | 183 $F{\cdot}g^{-1}$ | 1 $A{\cdot}g^{-1}$ | 81.5% (200 cycles) | 65 |
| $SnO_2$-$V_2O_5$-CNT (One-pot hydrothermal) | 0.1 M KCl | 121.39 $F{\cdot}g^{-1}$ | 100 $mV{\cdot}s^{-1}$ | 85.8% (100 cycles) | 66 |

Large scale production, however, can perhaps be considered for the most stable phase of $V_2O_5$ with its optimum performance as a supercapacitor. Thus, looking into the economic prospect, materials involving sol-gel route synthesis [60, 63, 67] and electrospinning [53] may be adopted for the growth of various $V_2O_5$.

## 5. Electrochemical properties of $VO_x$:

Other than the stoichiometric oxides of V, there are also reports of electrochemical applications of multi-valent vanadium oxides ($VO_x$). The V-O phase diagram comprises mixed-valence oxides comprehending two oxidation states, namely, $V_3O_7$, $V_6O_{13}$, $V_8O_{15}$, $V_7O_{13}$, $V_6O_{11}$, among others [68]. It permits conversion between oxides of different stoichiometry. Huang *et al.* [69] reported a high areal capacitance of 1.31 $F \cdot cm^{-2}$ from $VO_x$ functionalized by a carbon nanowire array. $V_3O_7$ was reported to get converted to $V_6O_{13}$ at the lowest potential of -0.6 V and $V_2O_5$ at the highest potential of 0.2 V. Zhao *et al*. [70] prepared NC-coated nest-like $V_3O_7$ which showed the specific capacity of 660.63 $F \cdot g^{-1}$ at 0.5 $A \cdot g^{-1}$ (figure 6a), a significantly higher than that of $V_3O_7$ (362.63 $F \cdot g^{-1}$ ). NC-$V_3O_7$ exhibited 80.47% of the initial capacitance at 10 $A \cdot g^{-1}$ after 4000 cycles, which is 23.16% higher than that of $V_3O_7$ (figure 6b).

$V_6O_{13}$ has stimulated extensive attention owing to its high specific capacitance and decent cycle ability for Li batteries. However, there are very few reports on its SC performance. $V_6O_{13}$ is known as a mixed-valence oxide as it exists between the $V^{4+}$ and $V^{5+}$ oxidation states with 2:1 ratio [71], which increases the electronic conductivity of the material. Zhai *et al*. [72] reported $V_6O_{13}$ as well as sulfur-doped, oxygen-deficient $V_6O_{13-x}$ as an anode electrode.

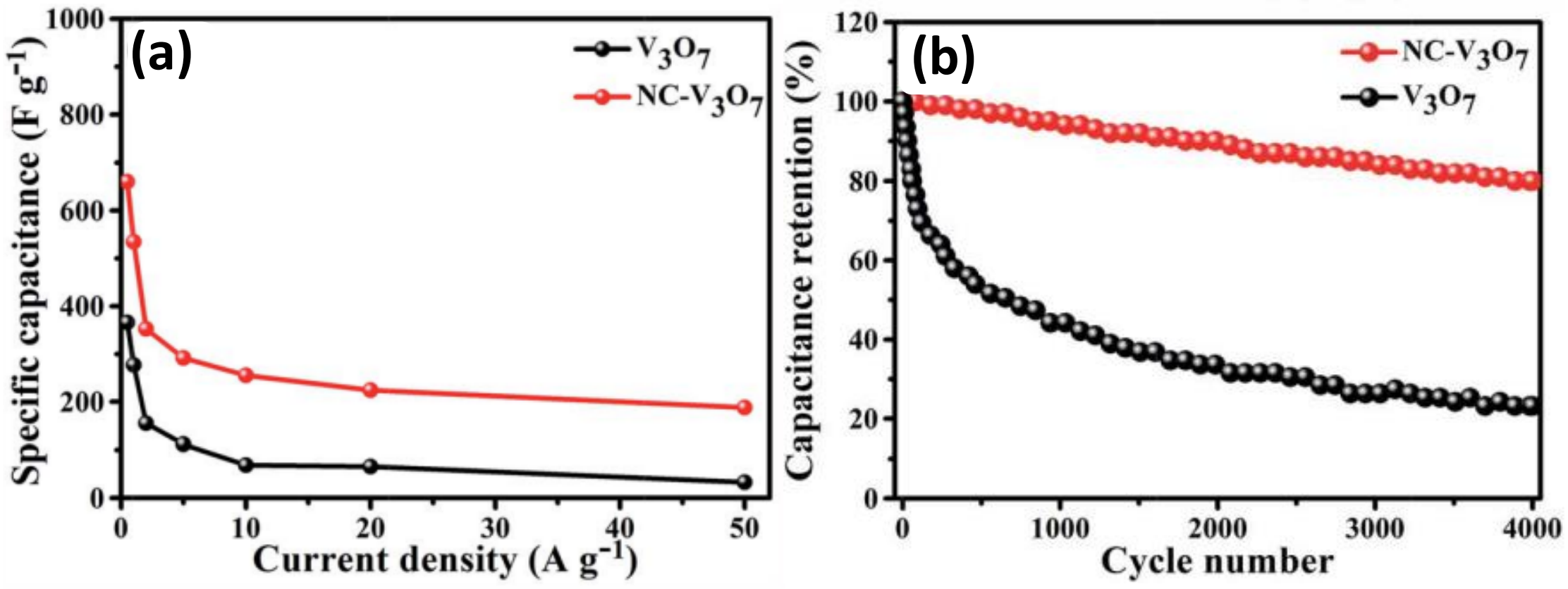

Figure 6. (a) Specific capacitance of $V_3O_7$ and NC-$V_3O_7$ as a function of current density and (b) cycling performance of $V_3O_7$ and NC-$V_3O_7$ at a current density of 10 A g$^{-1}$ [*Reprinted with permission from Ref. 69, Copyright 2019 @ Royal Society of Chemistry*].

$V_6O_{13-x}$ provides a capacitance of 1353 F·g$^{-1}$ at a current density of 1.9 A·g$^{-1}$ and outstanding capacitance retention of 92.3% after 10000 cycles. Pang *et al*. [73] reported the high electrochemical performance of 3D microflower structure of $V_4O_9$ supercapacitors (~392 F·g$^{-1}$).

## 6. Conclusion and outlook:

This brief review deals with the electrochemical performances of vanadium oxides and its composites as supercapacitors. Vanadium oxides have attracted tremendous attention in electrochemistry due to their multi-valency, low cost and abundant sources on earth. However, poor electrical conductivity, structural instability, poor specific capacitance and low energy density limit their practical applications. The scientific community is working towards removing the obstacles by morphology engineering (increase the specific surface area), doping with a heteroatom (reduce hydrophobicity and structural stability), combined with carbon-based materials and/or conducting polymers (increasing conductivity) and so on. The structure designing increases the specific surface area and offer more active sites which intern increases the contact between the material and electrolyte, generates more redox reactions, and enhances the electrochemical performance of the materials. It has been observed from the previous studies that nanostructures with porosity are the best choice for increasing supercapacitive performance. On the other hand, combining with composites increases the specific capacity, cyclic stability, and finally energy and power density. Comments are also made for the commercial viability in terms of large area synthesis of a stable phase of vanadium oxide with optimized supercapacitance properties. Finding a proper composite material for a specific vanadium oxide is still a challenge for future development. Finally, as vanadium oxides are prone to change the oxidization state, in-situ characterization techniques are likely to be carried out during the electrochemical processes. Most of the vanadium oxides also undergo electrical, magnetic, and/or structural transition with minimal change in the electric field, temperature, or pressure. Therefore, to understand the change in phases and its role in the electrochemical process, advanced in-situ characterization techniques should be incorporated.